# Quantum Degenerate Exciton-Polaritons in Thermal Equilibrium

Hui Deng, David Press, Stephan Götzinger, and Glenn S. Solomon

*Quantum Entanglement Project, ICORP, JST, Edward L. Ginzton Laboratory, Stanford University, Stanford, California 94305, USA*

Rudolf Hey and Klaus H. Ploog

*Paul-Drude-Institut für Festkörperelektronik, Hausvogteiplatz 5-7, D-10117 Berlin, Germany*

Yoshihisa Yamamoto

*Quantum Entanglement Project, SORST, JST, Edward L. Ginzton Laboratory, Stanford University, Stanford, California 94305, USA
and National Institute of Informatics, Hitotsubashi, Chiyoda-ku, Tokyo 101-8430, Japan*

We study the momentum distribution and relaxation dynamics of semiconductor microcavity polaritons by angle-resolved and time-resolved spectroscopy. Above a critical pump level, the thermalization time of polaritons at positive detunings becomes shorter than their lifetime, and the polaritons form a quantum degenerate Bose-Einstein distribution in thermal equilibrium with the lattice.

Bose-Einstein condensation (BEC) has been of intense interest to the physics community for decades [1–5]. While atom BEC has been demonstrated since 1995 in various species of atomic gases, no analogue has been established in solid state systems. The outstanding problem for solid state BEC is to have an equilibrium system with high enough density. Earlier works in $Cu_2O$ excitons showed thermal equilibrium but no quantum degeneracy [6,7]. More recent works on quantum-well excitons showed indirect evidence of quantum degeneracy, but thermal equilibrium could not be inferred [8–11]. Solid state polariton systems are interesting because they have an effective mass 8 orders of magnitude smaller than the hydrogen atom mass, and 4 orders of magnitude smaller than the exciton mass. Thus the critical temperature of polariton phase transitions range from 1 K to above room temperature. Quantum degeneracy has been demonstrated by many groups in recent years [12–16], yet thermal equilibrium is never established. In the current work, we obtained for the first time clear and direct evidence of simultaneous thermal equilibrium and quantum degeneracy.

In a semiconductor microcavity with embedded quantum wells (QWs), when the confined cavity photon modes strongly couple to the QW excitons, new eigenmodes are formed called the polaritons [17]. As quasiparticles in semiconductors, polaritons have relatively short lifetimes, thus it is generally difficult to cool hot polaritons to the lattice temperature before they decay. On the lower energy branch, polaritons change from excitonlike lower polaritons (ELPs) at large in-plane wave number $k$ to half-exciton half-photon lower polaritons (LPs) at $k \approx 0$. Correspondingly, their lifetime decreases by 2 orders of magnitude and their energy density of states decreases by 4 orders of magnitude. Hence an energy relaxation bottleneck is commonly observed [18–20] at low densities where spontaneous linear phonon-LP scattering is the dominant, yet insufficient, cooling mechanism. At higher densities, however, when the quantum degeneracy condition of $N_{LP} \gtrsim 1$ is fulfilled, bosonic final state stimulation greatly enhances both the nonlinear LP-LP scattering and the linear LP-phonon scattering [21–23]. Recently, a degenerate Bose-Einstein distribution (BED) of LPs has been observed [24], but the LP temperature was $T_{LP} \approx 100$ K, much higher than the lattice temperature $T_{lat} \approx 4.2$ K. This suggests that although LP-LP scattering establishes quasiequilibrium among LPs, cooling by the phonon bath is still slower than the decay of the LPs.

Fortunately the lifetime and thermalization time of the polaritons can be controlled by adjusting the detuning $\Delta$ of photon energy relative to the exciton energy at $k = 0$. With increasing positive $\Delta$, the excitonic fraction of a LP $\alpha_{exc} = \frac{4\Omega^2}{[\sqrt{\Delta^2+4\Omega^2}-\Delta]^2+4\Omega^2}$ increases, while the energy density of states $\rho(E) \propto (1 - \alpha_{exc})^{-1}$ also increases (where $2\Omega$ is the upper-lower polariton splitting). Hence the LP lifetime $\tau_{LP} \approx \tau_{cav}/(1 - \alpha_{exc})$ becomes longer, and LP-phonon scattering rates [$\propto \alpha_{exc}\rho(E)$] and LP-LP scattering rates [$\propto \alpha_{exc}^2\rho(E)$] become larger. At moderate positive detunings, we observe a quantum degenerate polariton gas in thermal equilibrium with the lattice. This is confirmed by measuring both the LP's thermalization time vs lifetime, and time-resolved momentum distributions of the LPs. At very large positive detunings, the observation of cooperative effects is eventually prohibited by the higher critical density, stronger exciton localization, and shorter dephasing time of ELPs.

The sample studied in this Letter consists of a $\lambda/2$ GaAs cavity sandwiched between $Ga_{0.865}Al_{0.135}As/AlAs$ distributed Bragg reflectors. Three stacks of QWs are placed at the central three antinodes of the microcavity, each stack consisting of four 6.8 nm-thick GaAs QWs separated by 2.7 nm-thick AlAs layers. The microcavity thickness is tapered to allow tuning of the cavity resonance from the center to the edge of the sample. We focus a linearly polarized picosecond mode-locked Ti:sapphire laser onto a 50 $\mu$m-diameter spot on the sample. The LP population per mode $N_{LP}(k)$ is measured directly by angle-resolved photon flux since there exists a one-to-one correspondence between the two via the $k$-dependent lifetime of the LPs. At an incidence angle of 50° from the sample growth direction, the laser resonantly pumps the corresponding ELP modes. The emitted light is collected with an angular resolution of 0.5° in air by an optical fiber which is in turn connected to an imaging spectrometer and/or a streak camera with a time resolution of 4 ps. Details of the experimental setup can be found in Ref [24].

In all experiments, we attach the sample to a copper cold finger kept at 4.2 K. Reflection and dispersion measurements show that the uncoupled exciton resonance is 1.597 eV, and the splitting between upper and lower polariton is $2\hbar\Omega = 14.4$ meV at zero detuning. We measured the LP emission intensity at $k = 0$ vs pump power $P$ and observed a quantum degeneracy threshold for detunings between $-5$ and $+10$ meV [Fig. 1(a)]. When the estimated LP number per pulse $I_{LP}(k = 0) \approx 1$ [25], $I_{LP}$ starts to increase nonlinearly with $P$, indicating the onset of stimulated scattering into the state. We define the operational threshold $P_{th}$ at the steepest slope of the input-output curve. $P_{th}$ ranges from 5 to 20 mW, corresponding to injected ELP densities $n_{QW} \approx 1$–$5 \times 10^9$ cm$^{-2}$ per QW. At the highest pump power of $P_{max} = 100$ mW, $n_{QW} \approx 2.5 \times 10^{10}$ cm$^{-2}$ per QW. The observed mean-field blueshift of the LP energy $\delta E_{LP}(k = 0)$ is less than 1 meV for all detunings and pump powers. At $\Delta = 6.7$ meV, $\delta E_{LP}(k = 0)$ is 0.2 to 0.3 meV from $P_{th}$ to $P_{max}$. Energy and momentum dispersion relations below and above threshold are measured to further confirm that the system stays in the strong coupling regime.

A prerequisite for reaching thermal equilibrium is that the LP's thermalization time must be faster than their lifetime. Shown in Fig. 1(b) is the time evolution of $N_{LP}(k = 0)$. To estimate the LP energy thermalization time vs lifetime, we model the system with two coupled modes, the $k \approx 0$ LP ground state and the hot-ELP reservoir:

$$\frac{\partial N_R}{\partial t} = P(t) - \frac{N_R}{\tau_R} - \frac{N_R}{\tau_{therm}}, \quad \frac{\partial N_0}{\partial t} = -\frac{N_0}{\tau_0} + \frac{N_R}{\tau_{therm}}, \quad (1)$$

where $N_R$ and $N_0$ are the ELP reservoir and LP ground state populations, respectively. $P(t)$ is an external pump represented by a Gaussian pulse, centered at $t = 0$ with a pulse width of 3 ps. $\tau_{therm}$ is the thermalization time from ELPs to the LP ground state. $\tau_R$ and $\tau_0$ are the lifetime of the ELPs and LP ground state, respectively. $\tau_0$ can be determined from the cavity lifetime of $\tau_{cav} = 2$ ps and the detuning-dependent photon fraction $\alpha_{cav}$ of the LP ground state: $\tau_0 = \tau_{cav}/\alpha_{cav}$. We use the normalized $N_0(t)$ to fit the experimental curve. Since the nonradiative lifetime of an exciton is on the order of 100 ps to 1 ns, $\tau_R$ is much longer than $\tau_0$ and $\tau_{therm}$. We found that it has little

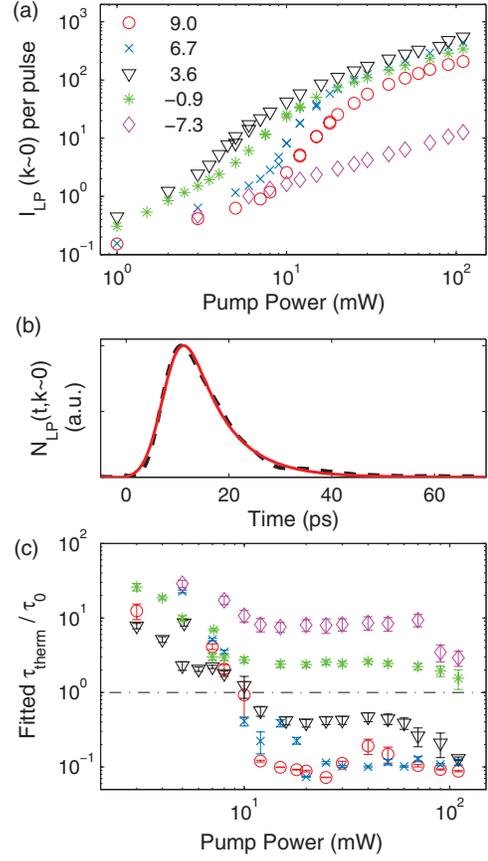

FIG. 1 (color). (a) LP number per pulse $I_{LP}(k \approx 0)$ [25] vs pump power at different detunings. No threshold is observed for $\Delta = -7.3$ meV. (b) Time evolution of ground state LP emission intensity $N_{LP}(k \approx 0)$. The dashed line is data taken for $\Delta = 6.7$ meV, $P = 20$ mW, deconvolved with the streak camera resolution of 4 ps. The solid line is a fit to Eq. (1). (c) Normalized thermalization time $\tau_{therm}/\tau_0$ vs pump power at different detunings. Below the dash-dotted line, $\tau_{therm}$ becomes shorter than $\tau_0$, so thermal equilibrium is expected. Detunings in meV are given in the figure legend.

influence on the fitting, hence we set $\tau_R$ to infinity for simplicity. $\tau_{therm}$ is the single fitting parameter. Despite the simplicity of the model, good agreement between data and the fitting curve is obtained as shown in Fig. 1(b).

In Fig. 1(c), we plot the normalized thermalization time $\tau_{therm}/\tau_0$. For all detunings there is a steep decrease of $\tau_{therm}$ near the quantum degeneracy threshold, reflecting the onset of stimulated scattering into the LP ground state. At a positive detuning of $\Delta = 6.7$ meV, $\tau_{therm}$ shortens from more than $20\tau_0$ below threshold to around $\tau_0$ near threshold, and saturates at about $0.1\tau_0$ well above threshold. We expect that thermal equilibrium with the phonon bath may therefore be established above threshold with a quantum degenerate LP population in the ground state. For negative detunings, $\tau_{therm}/\tau_0$ saturates above unity and the system is expected to stay in a nonequilibrium condition.

We then measure the time-resolved LP momentum distribution $N_{LP}(t, k)$ [25] and compare it to the classical

Maxwell-Boltzmann distribution (MBD) $N_{MB}(k)$ and the quantum mechanical Bose-Einstein distribution $N_{BE}(k)$. Referencing the LP band bottom as energy zero, we have

$$N_{MB}(k) = N_0 \exp\left(-\frac{E_{LP}(k)}{k_B T_{LP}}\right),$$
$$N_{BE}(k) = 1 \bigg/ \left[\exp\left(\frac{E_{LP}(k)}{k_B T_{LP}}\right)(1 + N_0^{-1}) - 1\right]. \quad (2)$$

The fitting parameters are the LP temperatures $T_{LP}$, and the LP population $N_0$ at $k = 0$. The normalized chemical potential $\alpha$ is defined by:

$$\alpha = -\mu/k_B T_{LP} = \ln(1 + N_0^{-1}). \quad (3)$$

At positive detunings $\Delta > 1$ meV, no bottleneck is observed at any pump level. Above threshold, from about 30 ps after the pump pulse, the experimental data are very well described by a BED, but not at all by a MBD, as shown in Fig. 2(a). From BED fitting, we obtain a $T_{LP}$ close to the lattice temperature $T_{lat}$, and $\alpha = 0.01$–0.5. Note that quantum degeneracy threshold $N_0 = 1$ is reached at $\alpha =$ ln2 $\approx 0.7$. The results indicate that LPs are well thermalized with the phonon bath, forming a quantum degenerate Bose-Einstein distribution at the lattice temperature. This is consistent with the observation that $\tau_{therm} \ll \tau_0$ at positive detunings above threshold.

At small detunings $|\Delta| < 1$ meV, although a bottleneck exists at most pump levels, the quantum degeneracy threshold is still achieved at $k \approx 0$. The momentum distribution in the region $|k| < 0.4$–1 $\mu m^{-1}$ agrees well with BED [Fig. 2(b)]. However, the fitted $T_{LP}$ is larger than $3T_{lat}$. This suggests that LPs with small $k$ can temporarily reach quasiequilibrium among themselves via efficient LP-LP scattering, yet decay out of the system before they can be sufficiently cooled by phonon emission. At large negative detunings $\Delta < -1$ meV, a strong bottleneck prevents the system from reaching thermal equilibrium [Fig. 2(c)].

In Fig. 3 we plot the time evolution of the fitted LP temperature $T_{LP}$ and normalized chemical potential $\alpha$. At 30–40 ps after the pump pulse injects hot ELPs, phonon-scattering cools the system to a lowest temperature of $T_{min} \approx T_{lat}$ for $\Delta = 6.7$ and 9.0 meV, with a fitted chemical potential $\alpha \approx 0.1$. The quantum degenerate LP gas remains in thermal equilibrium at $T_{lat}$ for a duration of about 20 ps. At the tail of the pulse, the LP population largely decays out of the system, $\mu$ increases to about $k_B T_{LP}$, and stimulated scattering diminishes. Since LPs at small $k$

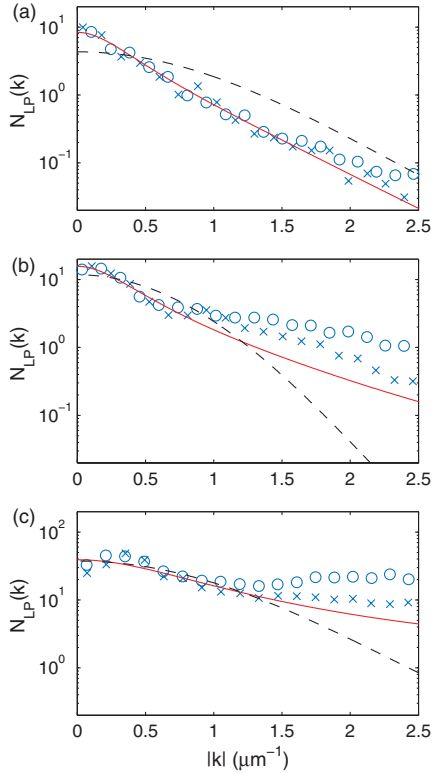

FIG. 2 (color online). Semilog plot of LP number per state $N(t, k)$ vs in-plane wave number $|k|$ for different detunings at a time when $T_{LP}$ reaches $T_{min}$: (a) $\Delta = 6.7$ meV, $P \approx 4P_{th}$; (b) $\Delta = -0.9$ meV, $P \approx 4P_{th}$; (c) $\Delta = -3.85$ meV, $P \approx 6P_{th}$. Crosses are data taken from the incidence direction of the pump, open circles are from the reflection direction, solid lines are fitting by BED, and dashed lines by MBD. Parameters for the curves are: (a) $T_{MB} = 4$ K, $T_{BE} = 4.4$ K, $\mu_{BE} = -0.04$ meV; (b) $T_{MB} = 4$ K, $T_{BE} = 8.1$ K, $\mu_{BE} = -0.13$ meV; (c) $T_{MB} = 8.5$ K, $T_{BE} = 182$ K, $\mu_{BE} = -0.4$ meV.

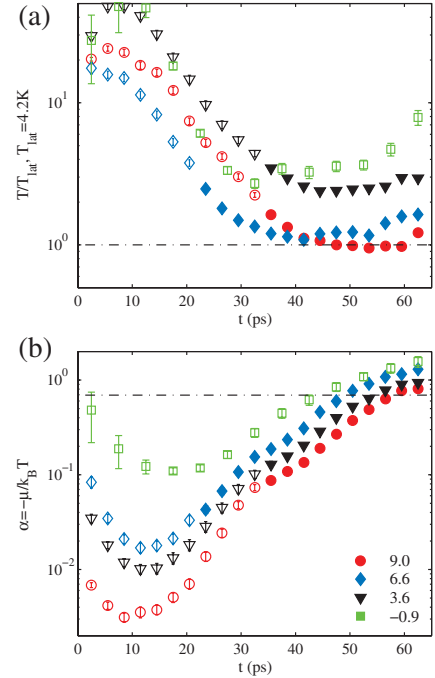

FIG. 3 (color). Time evolution of (a) normalized LP temperature $T_{LP}/T_{lat}$ and (b) normalized chemical potential $\alpha = -\mu/k_B T$, obtained from BED fitting of the LP momentum distribution, at $T_{lat} = 4.2$ K and $P \approx 3P_{th}$. $\Delta$ in meV is given in the legend. Closed symbols are $T_{LP}$ of thermalized LPs. Open symbols denote cases when the LPs are not fully thermalized; $T_{LP}$ does not correspond to an actual temperature and is plotted rather as a reference to help qualitatively understand the dynamics.

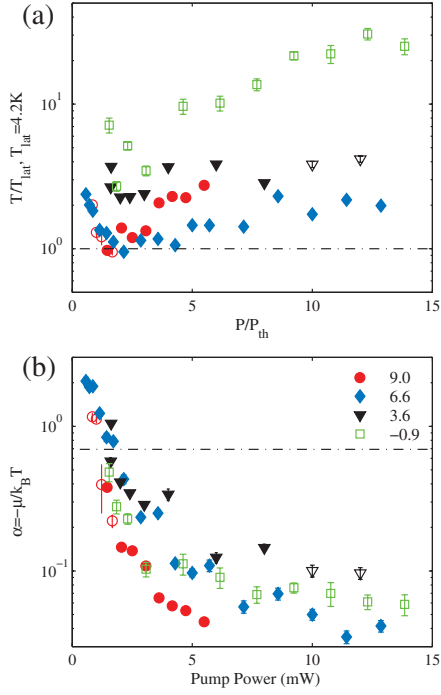

FIG. 4 (color). Pump dependence of (a) $T_{LP}/T_{lat}$ and (b) $\alpha = -\mu/k_B T$, obtained from BED fitting of the LP momentum distribution. Time $t$ is chosen as when $T_{LP}$ is at its minimum. $T_{lat} = 4.2$ K, $P \approx 3P_{th}$. $\Delta$ in meV is given in the legend. Closed (open) symbols denote cases when thermal equilibrium is (is not yet) established.

decay faster than those at larger $k$, the temperature of the system starts to rise. The smallest chemical potential $\alpha = 3 \times 10^{-3}$–$10^{-1}$ is found at 10–20 ps after the pump pulse when $N_{LP}(t)$ is at its peak. For smaller detunings, $T_{min}$ is higher than $T_{lat}$, so thermal equilibrium is not reached.

We choose a time 5–10 ps after the LP temperature reaches $T_{min}$, when the BED agrees very well with the data, and plot the pump power dependence of $T_{LP}$ and $\alpha$ in Fig. 4. As shown in Fig. 4(b), $\alpha$ decreases with increasing pump level, crossing the quantum degeneracy threshold $\alpha = 0.7$ at $P \approx P_{th}$. This is because the stimulated scattering into the LP ground state becomes prominent when the quantum degeneracy of LPs is established. In Fig. 4(a), $T_{LP}$ drops to a minimum value at $P = 2$–$4P_{th}$. At approximately the same pump level, the thermalization time $\tau_{therm}$ reaches its minimum [Fig. 1(c)]. At high pump levels, we notice that $T_{LP}$ increases, possibly because $T_{lat}$ increases due to insufficient heat dissipation of the sample to the copper cold finger.

In summary, increasing the detuning of the polariton system leads to a longer ground state lifetime $\tau_0$ and faster thermalization time $\tau_{therm}$. The system is tuned from a nonequilibrium state at negative detunings, to a quasithermal equilibrium of polaritons with high effective temperature near zero detuning, to a quantum degenerate thermal equilibrium at moderate positive detunings. For the positive detuning case, $\tau_{therm}$ is only one tenth of $\tau_0$, the LPs remain in thermal equilibrium with the phonon bath for a period of about 20 ps with a chemical potential $\mu \sim 0.1 k_B T$, well below the quantum degeneracy limit of $0.7 k_B T$.